\documentclass{aa}
\usepackage{epsf}

\begin{document}
\sloppypar

   \thesaurus{06     
              (02.01.2;  
               02.09.1;  
	       08.02.3;  
	       08.14.1;  
               13.25.3;  
               13.25.5)} 
   \title{High frequencies in the power spectrum of Cyg X-1 in the
   hard and soft spectral states.}

   \author{M. Revnivtsev \inst{1,2}, M. Gilfanov \inst{2,1},
   E. Churazov \inst{2,1}}

   \offprints{revnivtsev@hea.iki.rssi.ru}

   \institute{Space Research Institute, Russian Academy of Sciences,
              Profsoyuznaya 84/32, 117810 Moscow, Russia 
	\and
	      Max-Planck-Institute f\"ur Astrophysik,
              Karl-Schwarzschild-Str. 1, D-85740 Garching bei M\"unchen,
              Germany,
            }
  \date{}

        \authorrunning{Revnivtsev, Gilfanov\& Churazov}
        \titlerunning{High frequencies in the power spectrum of Cyg
        X-1}
        
   \maketitle

   \begin{abstract}
We analyzed a large number of RXTE/PCA observations of Cyg X-1 in the
hard and soft spectral states with total exposure time of $\approx
190$ and $\approx 10$ ksec respectively and time resolution better
than $\approx 250~\mu$s in order to investigate its variability down
to a few milliseconds time scales. The model of modifications of the power
density spectra due to the dead time effect was tested using RXTE
observations of an extremely bright source -- Sco X--1. 
The results of these tests demonstrated, that although some problems
still remain, for sources like Cyg X-1 (which is an order of magnitude less
bright than Sco X-1) present knowledge of the instrument is sufficient to
reliably analyze the power spectra in the kHz frequency range.

In both spectral states of Cyg X-1 we detected statistically significant
variability up to $\sim$150--300 Hz with a fractional rms amplitude
above 100 Hz at the level of $\sim2-3\%$.
The power spectrum in the hard state shows a 
steepening at frequency of $\sim 40-80$ Hz with the power law slope
changing from $\sim1.7$ to $\sim 2.3-2.4$. In the soft state the slope
of power spectrum changes from $\sim1$ to $\sim 2$ at the frequency of
$\sim 15-20$ Hz without any evidence of further steepening up to $\sim
100-150$ Hz. These break frequencies represent the highest
characteristic frequencies detected in the power density spectrum of
Cyg X-1 so far.

   \end{abstract}

%

\section{Introduction}

The shortest timescales in the variability of the compact objects always was
a very interesting topic of X-ray astronomy. 
The efforts to measure the highest frequency in the
variability of Cyg X-1 can be traced back to rocket experiments in 70-ties
(e.g. \cite{rothschild74}, \cite{giles81}). It was found that Cyg X-1
demonstrates statistically significant variability of the X-ray flux at least
up to frequencies of $\sim$10 Hz. Since then Cyg X-1 was 
extensively observed by various satellite missions searching for
millisecond and submillisecond time scales variability (HEAO-1,
see e.g. \cite{meekins84,wen96}; EXOSAT, e.g. \cite{belloni_has90}; GINGA,
e.g. \cite{miyamoto_cyg}). At the present time the Rossi X-ray
Timing Explorer observatory 
(RXTE, \cite{rxte}) is the  most powerful experiment for the investigation
of the short time scale X-ray flux variations. It combines high telemetry rate
(up to 512 kbps),
large effective area of the detectors ($\sim$6400 cm$^2$ in the soft
X-ray band), high time resolution (down to
1$\mu$sec) and comparatively low deadtime distortions. The comparison of
the characteristics of several X-ray experiments can be found in e.g.
\cite{giles98}.  

The shape of the power density spectrum (PDS) of Cyg X-1 in its
low/hard spectral state is well established (in the range from $\sim10^{-3}$ to
$\sim$50--100 Hz) since EXOSAT observations
-- it can be roughly represented by a constant 
from $\sim 10^{-3}$ Hz up to some break frequency ($P\sim f^{0}$), then
it steepens to the 
power law with index $\sim$1 ($P\sim f^{-1}$) and then it breaks again
to the power law with the index $\sim$1.6--1.8, $P\sim f^{-1.6-1.8}$
(see e.g. RXTE results in \cite{nowak}). Short observations ($\sim$10
ksec) of Cyg X-1 by RXTE/PCA 
with high time resolution ($\sim4 \mu$s) showed the absence of significant
variability of Cyg X-1 at time scales smaller than $\sim$3 msec
(\cite{giles98}). 

In this paper we report the systematic study of very high frequency
variability of Cyg X-1 in X-rays using large number of RXTE observations (total
exposure time close to 190 ksec for the hard state and $\sim$10 ksec
for the soft state) and carefully treating the deadtime
effects.

\section{Observations, data reduction and deadtime corrections}

For our analysis we used the data from the Proportional Counter Array
(PCA) aboard RXTE. To study the source
variability in the hard spectral
state, we chose a compact group of observations of Cyg X-1 performed
on 15--17 Dec. 1996 (proposal P10236) with total exposure time of
$\sim $80 ksec  and 56 available observations of Cyg X-1 from the proposal 
P30157 performed between Dec. 1997 and Dec. 1998 with total 
exposure time of $\approx$110 ksec. The soft spectral state data were  
accumulated on 4--18 June 1996 data (proposal P10512, see also
\cite{cui_cyg}) with total exposure 
time of $\approx$10 ksec.  All observations were
performed with  5 PCUs switched on.

The power density spectra were calculated from the light curves with
the bin duration equal to the intrinsic time resolution of the data
($2^{-13}$s$\approx 122\mu$s in proposals  P10236 and P10512 and
$2^{-12}$s$\approx 244\mu$s in proposal 
P30157). Some deviations from the noise level predicted by
the PCA dead time models are observed in the power spectra constructed
in the PCA energy sub-bands with the amplitude of $\sim10^{-6}$
(rms/mean)$^2$/Hz. We therefore followed recommendations of the RXTE
GOF and analyzed  the light curves in the total PCA energy band, where
these deviations seem to be considerably less significant (William
Zhang, private communication).

The power spectra calculated for individual time series were
normalized to units of squared fractional rms (\cite{miyamoto91})
$$
P_j=\left(\frac{2|a_j|^2}{N_{\gamma}}-2\right)~\frac{1}{R_{\rm tot}}
$$
and averaged. In the above formula $a_j$ is the Fourier
amplitude at the frequency $f_j$; $N_{\gamma}$-- is
total observed number of counts in the time series; $R_{\rm tot}$ --
averaged observed count rate for the time series. Note that the
ideal Poissonian noise component is already subtracted, therefore
$P_j=0$ for a Poissonian distribution of counts in the time series.

In a real detector various effects could lead to modification of the
Poissonian noise level by an amount $\Delta P$. Two major effects
contribute to $\Delta P$ in the case of PCA detector:
$$
\Delta P=\Delta P_{\rm dt}+\Delta P_{\rm vle}
$$ 
where $\Delta P_{\rm dt}$ is a modification of the noise level due to
counts dead time, $\Delta P_{\rm vle}$ -- additional noise component
resulting from  vetoing of the PCA detectors by the Very Large
Events (VLE). 

With an accuracy sufficient for our purpose the PCA dead time caused
by the incident photons can be described as a non-paralizable process
\footnote{However, there is small paralizable component in the total
  instrument deadtime (\cite{jahoda_dt1}). The value of paralizable deadtime
  in principle can slightly depend on the hardness of the source} 
(\cite{jahoda_dt1}), for which modification of the noise level was
derived by \cite{vihl94} (see also \cite{zhang_dt_95} for the 
expression accounting for finite length of the time series). We
shall use below the Eq. (A4) from \cite{vihl94} transformed to
the units of squared fractional rms:
\begin{eqnarray}
\Delta P_{\rm dt}(\tau_{\rm d},t_{\rm b},R) =-4 \cdot \frac{1}{R}
\cdot \sin(2\pi f t_{\rm b}/2)^2 \times
\nonumber \\
\sum_{k=-k_m}^{k=k_m} X[2\pi (f +k/t_{\rm b})]/(\pi (f t_{\rm b} +k))^2 
\end{eqnarray}
where
\begin{eqnarray}
X[f]=\frac{R'^2[1- \cos(2\pi f \tau_{\rm d})]+R'2\pi f \sin(2\pi f
\tau_{\rm d})}{R'^2[1- \cos(2\pi f \tau_{\rm d})]^2+[R' \sin(2\pi f
\tau_{\rm d})+2\pi f]^2};\nonumber \\
R'=\frac{R}{1-R\tau_{\rm d}}~~~~~~~~~~~~~~~~~~~~~~~~~~~~~~~~~~~~~~~~~~~~~~~~~~~~~
\nonumber 
\end{eqnarray}
Here $R$ is the observed count rate, $\tau_d$ is the photon dead time,
$t_{\rm b}$ is the bin duration, $f$ is the frequency. For the
practical purposes $k_m$ of 
the order of 10 is sufficient. This equation (for $k_m=\infty$) is valid for
an infinitely long time sequence. One can use more complicated expression
(see Eq. (44) from \cite{zhang_dt_95}) if the analyzed time
sequence contains only small number of bins. 

To the first approximation  the Very Large Events lead to appearance of
an additional (positive) component (\cite{zhang_dt_96})
\begin{eqnarray}
\Delta P_{\rm vle}(f,\tau_{\rm vle},R_{\rm vle}) = ~2{ R}_{\rm vle} \tau_{\rm vle}^2~ 
      \left [ {{\sin \left( \pi \tau_{\rm vle} f \right )} \over {\pi
 \tau_{\rm vle} f }}
      \right ]^2
\end{eqnarray}
where  $\tau_{\rm vle}$ -- the duration of the VLE window, $R_{\rm
vle}$ - the VLE count rate in one PCU. Note that the above expression
does not include the binning and sampling effects and is valid in the
limit $R_{\rm vle} \tau_{\rm vle}<<1$. 

Finally, the total noise component model in the first approximation would
be: 
$$
\Delta P=\frac{1}{N_{\rm pcu}}
\left[ 
\Delta P_{\rm dt}\left( \tau_{\rm d}, t_{\rm b}, R_{\rm pcu}\mu_{\rm vle} \right)
\mu_{\rm vle}
+\Delta P_{\rm  vle}\left(\tau_{\rm vle}, R_{\rm vle}  \right)
\right] 
$$
$$
\mu_{\rm vle}=\frac{1}{1-R_{\rm vle}\tau_{\rm vle}}
$$
where $R_{\rm pcu}$ - the {\em total observed} count rate in one PCU,
 $N_{\rm pcu}$ --
number of PCUs. The 
$\Delta P_{\rm dt}\left( \tau_{\rm d}, t_{\rm b}, R \right)$ and $\Delta P_{\rm
 vle}\left(\tau_{\rm vle}, R_{\rm vle}  \right)$ are given by the
Eq. (1) and (2) respectively. The factor $1/N_{\rm pcu}$ accounts for
the fact that the count streams from $N_{\rm pcu}$ independent units were
merged together. The factor $\mu_{\rm vle}$ accounts for reduction of the
observed count rate due to Very Large Events. This formula is valid under
the following conditions: $R_{\rm vle}<<R_{\rm pcu}$, $R_{\rm vle} \tau_{\rm 
  vle}<<1$ and $\tau_{\rm vle}>>\tau_{\rm dt}~R_{\rm pcu}~(\tau_{\rm
dt}+1/R_{\rm  pcu})$. 
 It is important to stress out that in the adopted dead time model
$R_{\rm pcu}$ is the 
the {\em total observed } count rate of the events causing the
dead time $\tau_d$ (as opposite to the observed count rate in the
analyzed time series). In the case of PCA detector it should include
total Good Xenon rate, Propane and Remaining counts. We also note that
for sufficiently small count rates, $R\la$3--5 kcnts/s/PCU
dependence of $\Delta P_{\rm dt}\left( \tau_{\rm d}, t_{\rm b}, R \right)$ and,
therefore, of $\Delta P$ upon $R$ vanishes. The model has been tested
on a series of Monte-Carlo simulations and has proven to be
sufficiently accurate in the parameters range of interest.

Note that we did not include in our model an additional background
term (\cite{pca_team_timing_sign} or \cite{jernigan} ). For bright
sources ($\ga$1 kcnts/s/PCA) its contribution can be neglected.

\begin{table*}
\caption{The expected and best fit model parameters of the deadtime
correction for 3 observations of Sco X-1. \label{dt_params}}
\tabcolsep=0.18cm
\begin{tabular}{cccccccccccc}
\hline
Obs.ID.&$t_{\rm b}$, $\mu$s&\multicolumn{2}{c}{$t_{\rm d}$, $\mu$s}&\multicolumn{3}{c}{$R_{\rm pcu}$,
kcnts/s}&\multicolumn{2}{c}{$\tau_{\rm vle}$, $\mu$s}&\multicolumn{2}{c}{$R_{\rm vle}$, cnts/s}&$\chi^2$/dof\\
&&\multicolumn{2}{c}{---------}&\multicolumn{3}{c}{-----------------------}&\multicolumn{2}{c}{---------}&\multicolumn{2}{c}{---------}\\
&&exp.&fit&Tot$^b$.&GX$^c$&fit&exp.$^d$&fit&exp.$^e$&fit&\\
\hline\\
\#1&15.25879&8--10$^a$&$8.5\pm0.1$&29.2&20.8&$17.6\pm0.3$&150&$152\pm1$&265&$274\pm2$&152/151\\
\#2&15.25879&8--10$^a$&$8.7\pm0.1$&19.6&15.1&$10.0\pm0.5$&60&$76\pm2$&239&$244\pm13$&180/157\\
\#3&122.0703&8--10$^a$&8.7(fixed) &23.6&18.3&$14$(fixed)&&76.0(fixed)&157&$162\pm6$&43/46\\
\hline
\end{tabular}
\\
\begin{list}{}
\item{$^a$} -- e.g. \cite{jahoda_dt1,jahoda_dt2,inflight}
\item{$^b$} -- measured total count rate (Good Xenon, Propane Counts
and Remaining Counts) per PCU
\item{$^c$} -- measured Good Xenon count rate per PCU
\item{$^d$} -- preflight values for all PCA, see \cite{pca_hk}
\item{$^e$} -- measured count rate (HouseKeeping data)
\end{list}
\end{table*}

\begin{figure}
\epsfxsize=8.5cm
\epsffile[20 180 570 700]{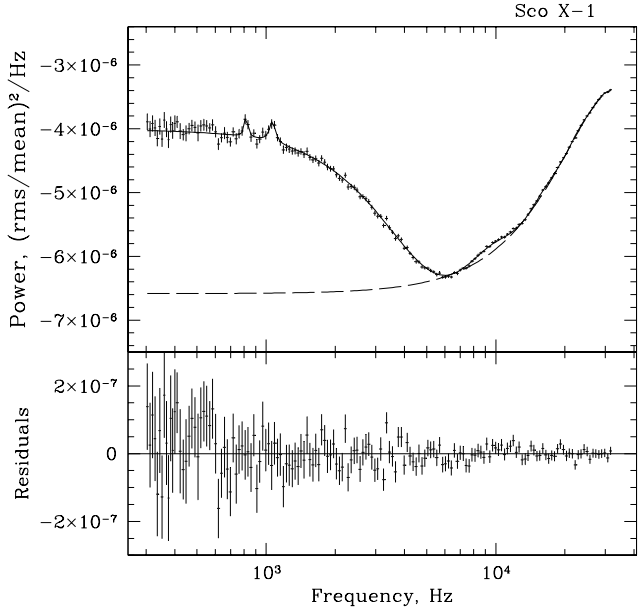}
\caption{The power spectrum of Sco X-1 from the observation \#1
performed with medium VLE window. The solid line shows the best fit
model consisting of the noise level model $\Delta P$ and two kHz
QPOs (see text). The dashed line shows the noise level component
$\Delta P_{dt}$ due to photon dead time. The lower panel shows the
residuals data--model. 
\label{scox1_ds2}}
\end{figure}

\begin{figure}
\epsfxsize=8.5cm
\epsffile[20 180 570 700]{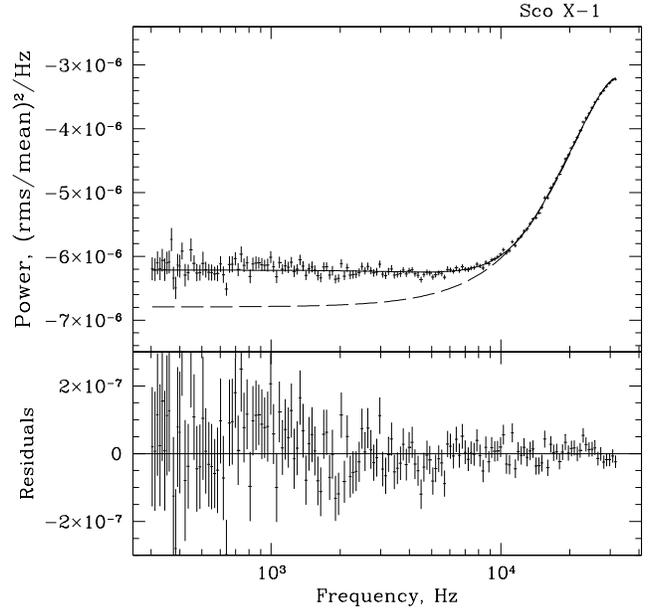}
\caption{The same as Fig.\ref{scox1_ds2} but for observation \#2,
performed with short VLE window.
\label{scox1_ds1}}
\end{figure}

\begin{figure}
\epsfxsize=8.5cm
\epsffile[20 180 570 700]{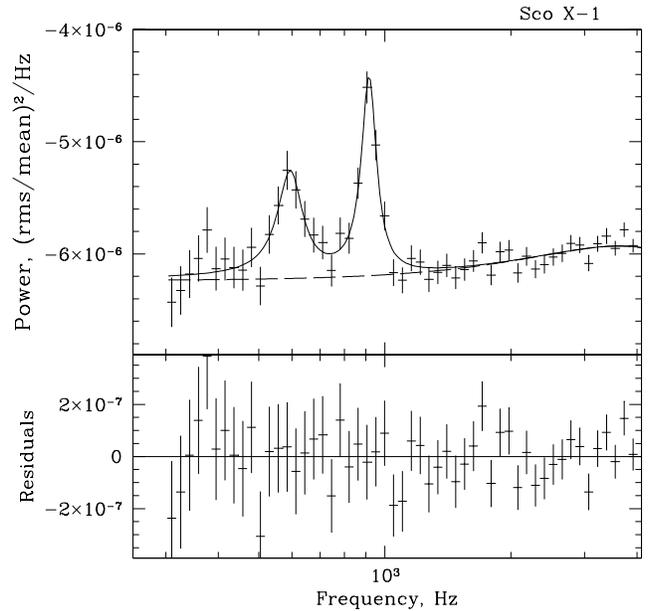}
\caption{The same as Fig.\ref{scox1_ds2} but for observation \#3,
performed with short VLE window and lower time resolution, $\approx
122 ~\mu$s.
\label{scox1_qpos}} 
\end{figure}

To verify our model for the noise component we compared it with the data
of PCA observations of an extremely bright source Sco X-1 ($\sim$100~000
cnts/s/PCA) for which all flavors of the dead time distortions are
much more prominent than for Cygnus X--1. We used 3 different
observations of Sco X-1 -- 10059-01-01-00 
(Feb. 14, 1996; hereafter \#1), 10059-01-03-00 (Feb. 19, 1996; \#2) and
10056-01-02-02 (May 26, 1996; \#3), having different value of the VLE
window $\tau_{\rm vle}$ and different time resolution. 
Ideally, once the values of the dead time $\tau_{\rm d}$ and VLE window
$\tau_{\rm vle}$ are calibrated, the model should reproduce the noise
level with count rates $R_{\rm pcu}$ and $R_{\rm vle}$ set to the measured
values. However, as we discuss below, this is not strictly the case. 
We therefore followed the  approach adopted by \cite{jernigan} and
fitted the power spectra at high frequencies, $f>300$ Hz, with the
model for the noise component plus the model for the source component
leaving the noise model parameters free. The source contribution
in this particular case was modeled as a superposition of two
Lorentzians representing kHz QPOs.  
This approach  has a disadvantage
that it requires a priory assumptions about the shape of the power
spectrum of the source. In particular, if the source has very weak and
very flat power spectrum component it might be not detected by our
procedure.

\begin{figure}
\epsfxsize=8.5cm
\epsffile[20 180 570 720]{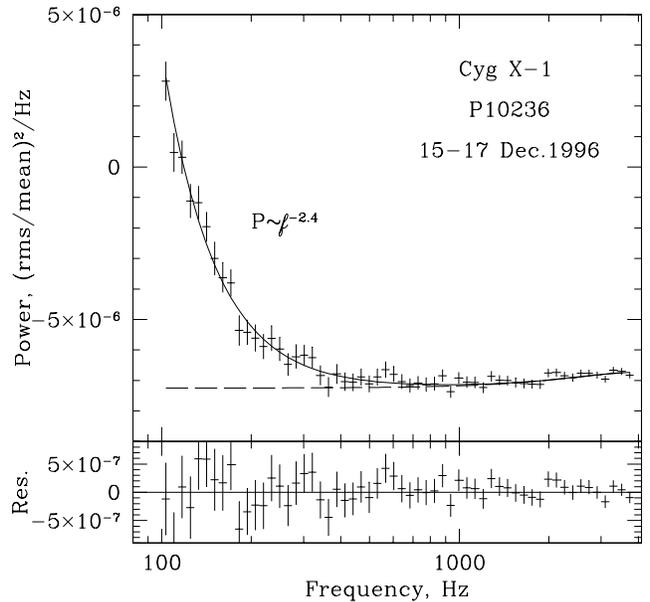}
\caption{The power spectrum of Cyg X-1 (observations 15--17 Dec.,
1996; proposal P10236). Total exposure time $\sim80$ksec. Dashed line
shows the instrumental noise level. \label{P10236_noise}}
\end{figure}

\begin{figure}
\epsfxsize=8.5cm
\epsffile[20 180 570 720]{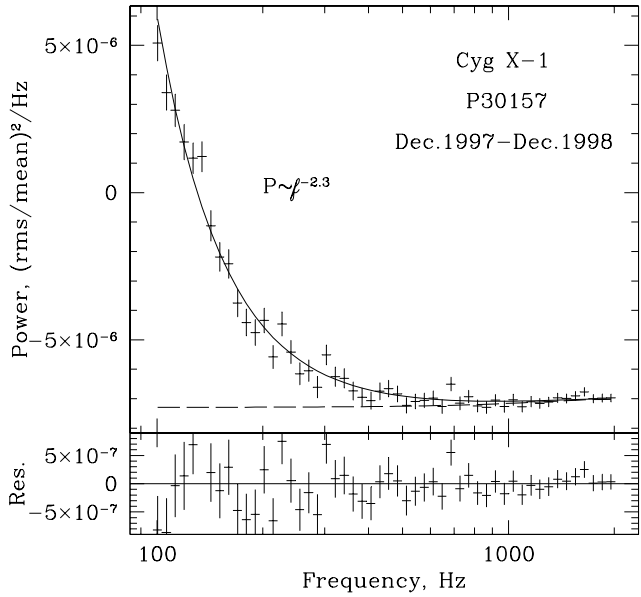}
\caption{The same as Fig. \ref{P10236_noise} but for observations from
proposal P30157 (Dec. 1997--Dec. 1998, hard spectral state). Total
exposure $\sim$110 ksec. 
\label{P30157_noise}}
\end{figure}

\begin{figure}
\epsfxsize=8.5cm
\epsffile[20 180 570 720]{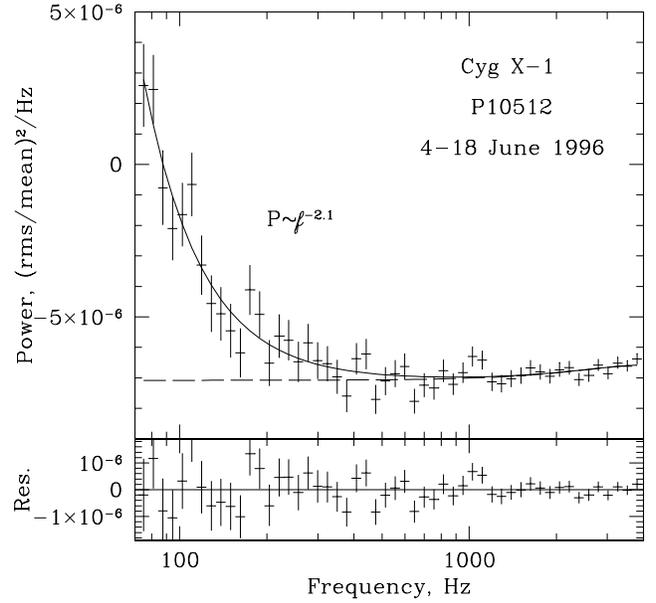}
\caption{The same as Fig. \ref{P10236_noise} but for 
observations from proposal P10512 (4-18 June, 1996, soft spectral
state). Exposure $\sim$10 ksec.
\label{soft_noise}}
\end{figure}

In Fig. \ref{scox1_ds2}, \ref{scox1_ds1} and \ref{scox1_qpos} we
present the power spectra of Sco 
X-1 in observations \#1 \#2 and \#3 along with the best fit noise
models and residuals. The best fit parameters and their expected values
are given in Table \ref{dt_params}.
For observation \#3 we fixed parameters $t_{\rm d},R_{\rm pcu}$
and $\tau_{\rm vle}$ because the limited frequency
range of the power spectrum ($<$4096 Hz) does not allow us to constrain
their values from the fit.
From Fig. \ref{scox1_ds2}, \ref{scox1_ds1} and
\ref{scox1_qpos} one can see that with appropriate tuning of the
parameters the adopted noise model is capable to reproduce the
observed shape of the noise component.

In general the best fit parameters of the noise level model
(Table \ref{dt_params}) look reasonable and are close to the expected 
values. The best fit value of $\tau_{\rm vle}$, in observation
\#2 is by $\approx 30\%$ higher than the preflight value. It is
possible, however, that the duration of the VLE window differs from
the preflight values (Alan Smale, private  communication).
The most apparent difference is a large, by a factor of $\sim 2$, discrepancy
between the best fit and observed total count rate. But we should note
that the observed GoodXenon count rate is much closer to the best fit
$R_{\rm pcu}$ value (see Table \ref{dt_params}). The reason of
this discrepancy is not clear, it might be due to unaccounted
processes in the PCA detectors at high count rates.  We note however, that dependence of the noise component
(in units of squared fractional rms) on the total count rate is a
specific feature of a 
non-paralyzable dead time process at sufficiently high count rates. At
lower count rate ($R_{\rm pcu}\la$3000--5000 cnts/s) and in particular in
the case of Cyg X-1 ($R_{\rm pcu}\sim$1000-1500 cnts/s) this
dependence vanishes.

\section{Results}

We adopted the noise level model described in the previous section for
the subsequent analysis of the Cyg X-1 data. Since the Cyg X-1
observations were performed with time resolution coarser than $\approx
122 ~\mu$sec,  the photon dead time and duration of the VLE window are
not constrained by the data. We fixed them at the best
fit values determined from the Sco X-1 data: $t_{\rm d}=8.7\mu$s and
$\tau_{\rm vle}=76\mu$s. As mentioned above, at the count rates
typical for Cyg X-1, $R_{\rm pcu}\sim$1000--1500 cnts/s, the noise
level does not 
depend on the count rate. We therefore fixed the $R_{\rm pcu}$ parameter
at the averaged observed value. The only free parameter of the noise
level model was $R_{\rm vle}$. 
Similarly to the procedure applied to the Sco X-1 data, we included
the source component into the model.  The source component was
represented by a power law  $P\propto f^{-\alpha}$. We used the high
frequency part of the power spectra, from 100 Hz up to the Nyquist
frequency, for the fits.

\begin{figure}
\epsfxsize 8cm
\epsffile[32 185 550 720]{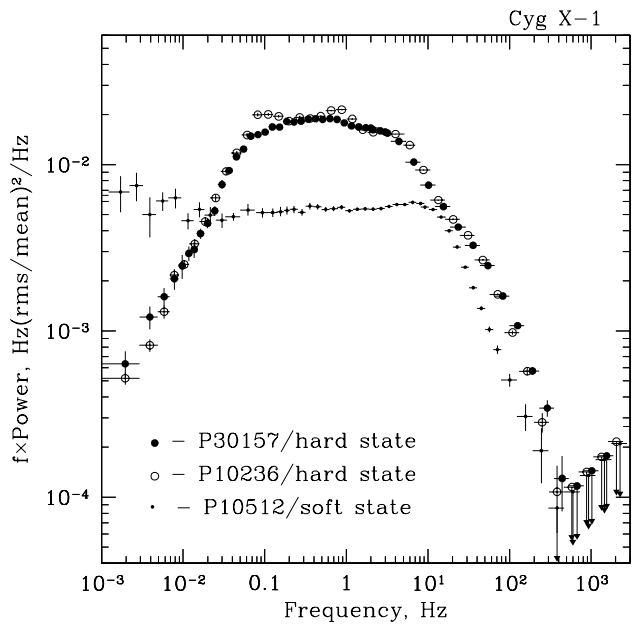}
\caption{The broad band ($10^{-3}-10^3$ Hz) power spectrum of Cyg X-1
in the hard and soft spectral states averaged over different
data sets used for the analysis. The power spectra were multiplied by the
frequency.
\label{broadband}} 
\end{figure}

The high frequency part of the observed power spectra of Cyg X-1
before subtraction of the noise level $\Delta P(f)$, the best fit
model and  the residuals of the data from the model are shown in
Fig. \ref{P10236_noise}, \ref{P30157_noise} and  \ref{soft_noise}. The
overall power spectra in a broad frequency range from 1 mHz to
$\approx 2$ kHz after subtraction of the best fit noise level model
are shown in Fig.\ref{broadband}. 
One can see that the  model describes the data
points reasonably well: $\chi^2\sim$38/55 dof for observations P10236 
(Fig. \ref{P10236_noise}), $\chi^2\sim$56/49 dof 
for observations P30157 (Fig. \ref{P30157_noise}) and
$\chi^2\sim$46/49 for observations P10512 (soft state,
Fig.\ref{soft_noise}).
The best fit values of the slope of the source 
power spectrum are: $\alpha=2.4\pm0.1$ for P10236, $\alpha=2.27\pm0.08$
for P30157 and $\alpha=2.1\pm0.2$ for P10512. The intrinsic
variability of the source has been statistically significantly
detected up to frequencies of $\sim$150--300 Hz
(Fig. \ref{P10236_noise}--\ref{soft_noise}) with
the fractional rms amplitude in frequency band $\sim$100-400 Hz of
$\approx3$\% (hard spectral state) and  $\approx$2\% (soft state). 

\begin{figure}[htb]
\epsfxsize 8cm
\epsffile[32 185 550 720]{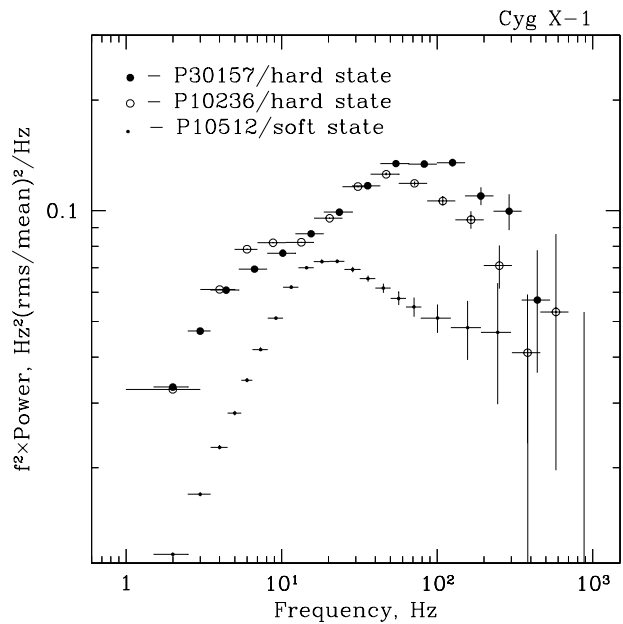}
\caption{The power density spectrum of Cyg X-1 at high
frequencies. Note, that the power spectrum has been multiplied by
$f^2$.
\label{outcome}}
\end{figure}

In order to increase statistics we averaged all the hard state data
increasing the total exposure time up to $\approx 190$ ksec. The power
law approximation of the averaged power spectrum above 100 Hz gives
the power law index of $\alpha=2.32\pm0.07$. The power law
approximation of the same data in the 20--60 Hz frequency range
results in $\alpha=1.66\pm0.01$, although the PDS clearly
has a more complicated shape than a power law (Fig.\ref{outcome}).  
We therefore conclude that we detected statistically significant
steepening of the power spectrum of Cyg X-1 in the hard state at
the frequency of $\sim$40--80 Hz. Note, that some indication of such  
steepening can be found in \cite{nowak}, although short duration of the 
used observations ($\sim 20$ ksec) was not sufficient to study the
high frequencies in detail. 
In order to illustrate the high frequency behaviour of the power
spectra, we plot in Fig.\ref{outcome} the power spectra multiplied by
$f^2$.  As it can be seen from Fig.\ref{outcome} the particular shape of
the rollover above 100 Hz is not significantly constrained by the
data. Assuming that power spectrum continues with the same slope to
the higher frequencies an observatory of the  EXTRA/LASTE class
(effective area $\sim10$ m$^2$, \cite{barret_extra}) would be able to detect
statistically significant signal up to several kHz in $\sim$tens of ksec
exposure time and help us to say something more about the exact shape of the 
rollover of the PDS.

The 2$\sigma$ upper limit on the amplitude of a possible QPO component
(Lorentz profile with quality $Q$) in the 500-2000 Hz frequency range
in the hard state power spectrum is $\sim$2\% for $Q=1$ and
$\sim$0.9\% for $Q=20$. These upper limits were obtained using the
observations from the propsal P10236 performed on 15-17 Dec. 1996
in which the shape of the power density at the lower frequency did not
vary significantly.
The upper limit on the continuum noise component at the high
frequency is somewhat more difficult to estimate. A statistical
($2\sigma$) error on the fractional rms in the 500-2000 Hz for the
sum of all hard state observations  is $\approx
1\%$. Assuming that the noise level model is exact the upper limit on
the fractional rms would coincide with the above number. However, as
we mentioned in the previous section a weak and very flat source power
spectrum component might be not detected given the procedure used to
determine the instrumental noise level.

\section{Summary}

We analized $\sim$190 ksec of observations of Cyg X-1 in the low/hard
state and reanalyzed $\sim$10 ksec of high/soft state observations.
We detected statistically significant variability
up to  frequencies of $\sim$300--400 Hz with the fractional rms
amplitude in the 100--400 Hz frequency range $\approx$3\% in the
hard  state and $\approx$2\% in the soft state.
The $2\sigma$ upper limits on the fractional rms amplitude of a QPO
feature with Lorentzian profile in the 500--2000 Hz frequency range
are $\sim$2\% for $Q\sim1$ and $\sim$0.9\% for $Q\sim20$ in the hard 
state and 2.5\% ($Q\sim1$) and 1.5\% ($Q\sim20$) in the soft state. Assuming
that the model for the instrumental noise level is exact, the
$2\sigma$ upper limits on the continuum variability in the 400--2000
Hz frequency range are $\sim$1\% and  $\sim1.7$\% for the hard and soft state
respectively.  

We detected statistically significant steepening of the power spectrum
in the hard state at frequencies $\sim40-80$ Hz. The power law
approximation to the spectrum at frequencies higher than $\sim$80--100
Hz gives the slope of $\alpha\sim$2.3--2.4.
In the soft state the slope
of power spectrum changes from $\sim1$ to $\sim 2$ at the frequency of
$\sim 15-20$ Hz without any evidence of further steepening up to the
frequency of $\sim 100-150$ Hz.

\section{Astrophysical implications of high frequency variability}

Thus the sensitivity and time resolution of the RXTE allow one to probe
interesting range of the variability time scales for accreting black holes.
Several characteristic frequencies fall approximately in this range. First
of all the frequency of rotation at the last marginally stable orbit for
the non-rotating (Schwarzschild) black hole is $f\sim200 \frac{10M_\odot}{M}$
Hz. For the maximally rotating Kerr black hole this frequency ranges from
$\sim$1200 Hz to $\sim150$ Hz for the co-rotating and counter--rotating orbits
respectively. If Keplerian rotation itself makes significant contribution to
the variability of the X--ray flux (e.g. due to the Doppler shift and boosting
of the emission from the ``hot spots'' in the disk) this difference should
be manifested in this frequency range (Sunyaev 1972). Secondly typical sound crossing time
of the disk height (in the case of optically thick geometrically thin
standard accretion disk) corresponds to the frequency of hundreds of Hz. If
strong turbulence is present 
in the accretion flow as anticipated by the most popular theoretical models
(see e.g. \cite{balbus} for review) then it may affect the variability at
higher 
frequencies. The dependence of the amplitude of variations on the frequency
is then a valuable tool to study the turbulence in the accretion flow. 
Thirdly light crossing time of the central region 
$t_{\rm lc}=3R_g/c\sim$ corresponds to the frequencies of the order of
kHz. The light crossing time for the region of the main energy release 
($R\sim$ 10--20 $R_g$)  corresponds to the frequencies of 300--500 Hz. The
character of the the variability may change drastically around this
frequency or (e.g. in the comptonization models involving multiple
scatterings of the photons) at a several times lower frequencies
(e.g. \cite{payne80}, \cite{st80}, \cite{pss83}, \cite{nowak_vaughan}).

The RXTE data (i) demonstrate that power specifically associated with any
of these characteristic frequencies is not large (see given above upper
limits on the narrow features on the PDS in the frequency range from
500--2000 Hz) -- i.e. there are no ``resonances'' at these frequencies, (ii)
limits the total power associated with high frequency variability
(from 400 Hz to 2000 Hz) to a percent level, (iii) show that in the
hard state of the 
source prominent, previously unknown, break in the PDS occurs at the
frequency of $\sim$ 70 Hz, (iv) extend the range of significantly detected
variations up to the frequencies of $\sim$200--300 Hz. The theoretical
models of the accretion onto the black holes are now to be tested against
these results. The study of the high frequency noise might become an
additional tool to distinguish black holes from weakly magnetized
neutron stars (Sunyaev \& Revnivtsev 2000). In the subsequent
publication (Revnivtsev et al. 2000, in preparation) we perform
similar analysis for the accreting neutron stars.

\begin{acknowledgements}
The authors thank Rashid Sunyaev for numerous discussions, William Zhang for
helpful discussions of PCA deadtime, 
Alan Smale and Keith Jahoda for the useful comments on the PCA operations. 
This research has made use of data obtained 
through the High Energy Astrophysics Science Archive Research Center
Online Service, provided by the NASA/Goddard Space Flight Center.
M.Revnivtsev acknowedges partial support by RFBR grant 00-15-96649. 
\end{acknowledgements}

\end{document}